\documentclass[12 pt,twoside,aps,prd,amsmath,amssymb,
tightenlines,showpacs,showkeys,eqsecnum]{revtex4-1}
\usepackage{mathptmx}
\usepackage{bm}
\usepackage{graphicx}
\usepackage{hyperref}
\renewcommand{\cal}{\mathcal}

\newcommand {\pr}{\partial}

\newcommand {\cG}{\cal G}
\newcommand {\cL}{\cal L}

\newcommand {\vf}{\varphi}

\def \myfigures #1#2#3#4#5#6#7#8
{\begin{figure}[ht]
    \begin{center}
        \includegraphics[width=#2 \textwidth]{#1.eps}
        \hfill
        \includegraphics[width=#6 \textwidth]{#5.eps}
        \parbox[t]{#4\textwidth}{\caption {#3}\label{#1}}
        \hfill
        \parbox[t]{#8\textwidth}{\caption {#7}\label{#5}}
    \end{center}
\end{figure} }

\def\myfigure #1#2#3#4
{\begin{figure}[ht]\begin{center}
\includegraphics[width=#2 \textwidth]{#1.eps}
\parbox[t]{#4\textwidth}{\caption{#3}\label{#1}}
\end{center}\end{figure}}

\date{\today}
\begin{document}
\title{Electromagnetic field with induced massive term: Case with scalar field}
\author{Yu.P. Rybakov, G.N. Shikin and Yu.A. Popov}
\affiliation{Department of Theoretical Physics\\
Peoples' Friendship University of Russia\\
117198 Moscow, Russia}\email{soliton4@mail.ru}

\author{Bijan Saha}
\affiliation{Laboratory of Information Technologies\\
Joint Institute for Nuclear Research, Dubna\\
141980 Dubna, Moscow region, Russia} \email{bijan@jinr.ru}
\homepage{http://bijansaha.narod.ru/}

\begin{abstract}
We consider an interacting system of massless scalar and
electromagnetic field, with the Lagrangian explicitly depending on
the electromagnetic potentials, i.e., interaction with broken gauge
invariance. The Lagrangian for interaction is chosen in such a way
that the electromagnetic field equation acquires an additional term,
which in some cases is proportional to the vector potential of the
electromagnetic field. This equation can be interpreted as the
equation of motion of photon with induced nonzero rest-mass. This
system of interacting fields is considered within the scope of
Bianchi type-I (BI) cosmological model. It is shown that, as a
result of interaction the electromagnetic field vanishes at $t \to
\infty$ and the isotropization process of the expansion takes place.
\end{abstract}



\maketitle

\bigskip


\section{Introduction}

The hypothesis of possible nonzero photon mass has long been
discussed in the literature \cite{gold,froome,taylor,broglie}. The
modern experimental data do not contradict this hypothesis
\cite{rosen,williams,crandal,chernikov,schaefer,fishbach,davis,lakes,luo}.
So it is interesting to consider some additional arguments for or
against this hypothesis. As one of such arguments can serve
experimental data of modern observational cosmology, which witnesses
the isotropy of the Universe. It is interesting to combine this fact
with the description of matter by means of system of interacting
fields including the electromagnetic one. It comes out that in a
number of cases consideration of such system in cosmology happens to
be equivalent to the photon mass. In this paper we consider one of
the simplest systems comprising with mass-less scalar and
electromagnetic  fields and study the influence of such interaction
on the expansion of the Universe in the asymptotic region.

\section{Basic equations and their general solutions}

We choose the Lagrangian of the interaction electromagnetic and
massless scalar fields within the framework of a BI cosmological
gravitational field in the form
\begin{equation}
{\cL} = \frac{R}{2\varkappa} - \frac{1}{4} F_{\mu\nu}F^{\mu\nu} +
\frac{1}{2} \vf_{,\eta} \vf^{,\eta}{\cG}, \quad {\cG}= (1 + \Phi(
I)),\ \label{lag}
\end{equation}
with $I = A_\mu A^\mu$.

We consider the BI metric in the form
\begin{equation}
ds^2 = e^{2\alpha} dt^2 - e^{2\beta_1} dx^2 - e^{2\beta_2} dy^2 -
e^{2\beta_3} dz^2. \label{BI}
\end{equation}
The metric functions $\alpha, \beta_1, \beta_2, \beta_3$ depend on
$t$ only and obey the coordinate condition
\begin{equation}
\alpha = \beta_1 + \beta_2 + \beta_3. \label{cc}
\end{equation}

Written in the form
\begin{equation}
R_{\mu}^{\nu} = -\varkappa \Bigl(T_{\mu}^{\nu} - \frac{1}{2}
\delta_{\mu}^{\nu} T \Bigr), \label{ee}
\end{equation}
the Einstein equations corresponding to the metric \eqref{BI} in
account of \eqref{cc} read
\begin{subequations}
\label{BID}
\begin{eqnarray}
e^{-2\alpha} \Bigl(\ddot \alpha - \dot \alpha^2 + \dot \beta_1^2 +
\dot \beta_2^2 + + \dot \beta_3^2 \Bigr) &=&  - \varkappa
\Bigl(T_{0}^{0} - \frac{1}{2} T \Bigr),\label{00}\\
e^{-2\alpha} \ddot \beta_1 &=&  - \varkappa
\Bigl(T_{1}^{1} - \frac{1}{2} T \Bigr),\label{11}\\
e^{-2\alpha} \ddot \beta_2 &=&  - \varkappa
\Bigl(T_{2}^{2} - \frac{1}{2} T \Bigr),\label{22}\\
e^{-2\alpha} \ddot \beta_3 &=&  - \varkappa \Bigl(T_{3}^{3} -
\frac{1}{2} T \Bigr),\label{33}
\end{eqnarray}
\end{subequations}

where over dot means differentiation with respect to $t$ and
$T_{\nu}^{\mu}$ is the energy-momentum tensor of the matter fields.

Variation of \eqref{lag} with respect to electromagnetic field gives
\begin{equation}
\frac{1}{\sqrt{-g}} \frac{\pr}{\pr x^\nu} \Bigl(\sqrt{-g}
F^{\mu\nu}\Bigr) - \Bigl(\vf_{,\nu} \vf^{,\nu}\Bigr) {\cG}_I A^{\mu}
= 0, \quad  {\cG}_I = \frac{d {\cG}}{dI}. \label{emf}
\end{equation}
The scalar field equation corresponding to the Lagrangian
\eqref{lag} has the form
\begin{equation}
\frac{1}{\sqrt{-g}} \frac{\pr}{\pr x^\mu} \Bigl(\sqrt{-g} g^{\mu\nu}
\vf_{,\nu} {\cG}\Bigr) = 0. \label{scf}
\end{equation}
The energy-momentum tensor of the interacting matters fields has the
form
\begin{eqnarray}
T_\mu^\nu &=& \Bigl[\vf_{,\mu}\vf^{,\nu}{\cG} - F_{\mu \eta}
F^{\nu\eta} + \bigl(\vf_{,\eta} \vf^{,\eta}\bigr) {\cG}_IA_{\mu}
A^{\nu}\Bigr],  \nonumber\\ &-& \delta_\mu^\nu \Bigl[ \frac{1}{2}
\vf_{,\eta} \vf^{,\eta} {\cG} - \frac{1}{4}
F_{\eta\rho}F^{\eta\rho}\Bigr]. \label{emt}
\end{eqnarray}
We consider the case when the electromagnetic and scalar fields are
the functions of $t$ only.  Taking this in mind we choose the vector
potential in the following way:
\begin{equation}
A_{\mu} = \bigl(0,\,A_1(t),\,A_2(t),\,A_3(t)\bigr). \label{vpot}
\end{equation}
In this case the electromagnetic field tensor $F^{\mu \nu}$ has
three non-vanishing components, namely
\begin{equation}
F_{01} = \dot A_1, \quad F_{02} = \dot A_2, \quad F_{03} = \dot A_3.
\label{emtensor}
\end{equation}
On account of \eqref{vpot} and \eqref{emtensor} we now have
\begin{eqnarray}
I &=&  - A_1^2 e^{-2\beta_1} - A_2^2
e^{-2\beta_2} - A_3^2 e^{-2\beta_3}, \label{eminv}\\
F_{\mu\nu}F^{\mu\nu} &=& - 2 e^{-2\alpha} \bigl(\dot A_1^2
e^{-2\beta_1} + \dot A_2^2  e^{-2\beta_2} + \dot A_3^2e^{-2\beta_3}
\bigr). \label{fmn}
\end{eqnarray}
Let us now solve the scalar field equation. Taking into account that
$\vf = \vf(t)$, from the scalar field equation one finds
\begin{equation}
\dot \vf = \frac{\vf_0}{{\cG}}, \qquad \Rightarrow
\vf_{,\mu}\vf^{,\mu} = \frac{\vf_0^2}{{\cG}^2} e^{-2\alpha}, \quad
\vf_0 = {\rm const.} \label{sfs}
\end{equation}
On account of \eqref{emtensor} and \eqref{sfs} for electromagnetic
field we find
\begin{subequations}
\label{emf123}
\begin{eqnarray}
\frac{d}{dt}\Bigl(\dot A_1 e^{-2\beta_1}\Bigr) - \vf_0^2 P_I
A_1 e^{-2\beta_1} &=& 0, \label{emf1}\\
\frac{d}{dt}\Bigl(\dot A_2 e^{-2\beta_2}\Bigr) - \vf_0^2 P_I
A_2 e^{-2\beta_2} &=& 0, \label{emf2}\\
\frac{d}{dt}\Bigl(\dot A_3 e^{-2\beta_3}\Bigr) - \vf_0^2 P_I A_3
e^{-2\beta_3}  &=& 0, \label{emf3}
\end{eqnarray}
\end{subequations}
where we set $P(I) = 1/{\cG(I)}$.

Finally, let us solve the Einstein equations. In doing so, let us
first write the nonzero components of the energy momentum tensor of
material fields. In view of \eqref{sfs} from \eqref{emt} we find
\begin{subequations}
\label{emtcomp}
\begin{eqnarray}
T_0^0 &=&\Bigl[\frac{\vf_0^2 P}{2}+\frac{1}{2} \Bigl(\dot A_1^2
e^{-2\beta_1} + \dot A_2^2 e^{-2\beta_2} + \dot
A_3^2 e^{-2\beta_3}\Bigr)\Bigr] e^{-2\alpha}, \label{00emt}\\
T_1^1 &=& \Bigl[- \frac{ \vf_0^2 P}{2} +\frac{1}{2} \Bigl(\dot A_1^2
e^{-2\beta_1} - \dot A_2^2 e^{-2\beta_2} - \dot A_3^2
e^{-2\beta_3}\Bigr)
+\vf_0^2 P_I  A_1^2 e^{-2\beta_1}\Bigr] e^{-2\alpha}, \label{11emt}\\
T_2^2 &=& \Bigl[- \frac{\vf_0^2 P}{2} +\frac{1}{2} \Bigl(\dot A_2^2
e^{-2\beta_2} - \dot A_3^2 e^{-2\beta_3} - \dot A_1^2
e^{-2\beta_1}\Bigr)
+\vf_0^2 P_I A_2^2 e^{- 2\beta_2}\Bigr]e^{-2\alpha}, \label{22emt}\\
T_3^3 &=& \Bigl[- \frac{\vf_0^2 P}{2}  +\frac{1}{2} \Bigl(\dot A_3^2
e^{-2\beta_3} - \dot A_1^2 e^{-2\beta_1} - \dot A_2^2
e^{-2\beta_2}\Bigr)
+ \vf_0^2 P_I  A_3^2 e^{- 2\beta_3}\Bigr]e^{-2\alpha}, \label{33emt}\\
T_2^1 &=& \Bigl(\dot A_1 \dot A_2 + \vf_0^2 P_I A_1
A_2\Bigr) e^{-2 \alpha -2\beta_1},\label{12emt}\\
T_3^2 &=& \Bigl(\dot A_2 \dot A_3 + \vf_0^2 P_I A_2
A_3\Bigr) e^{-2 \alpha -2\beta_2},\label{23emt}\\
T_1^3 &=& \Bigl(\dot A_3 \dot A_1 + \vf_0^2 P_I A_3 A_1\Bigr) e^{-2
\alpha -2\beta_3}.\label{31emt}
\end{eqnarray}
\end{subequations}
From \eqref{emtcomp} one also finds
\begin{equation}
T = \Bigl[ - \vf_0^2 P + \vf_0^2 P_I \Bigl(A_1^2 e^{-2\beta_1} +
A_2^2 e^{-2\beta_2} + A_3^2 e^{-2\beta_3}\Bigr)\Bigr] e^{-2\alpha} =
- \vf_0^2 \bigl[P + I P_I\bigr]e^{-2\alpha}. \label{T}
\end{equation}

The  triviality of off-diagonal components of the Einstein tensor
for BI metric leads to
\begin{equation}
T_2^1 = T_3^2 = T_1^3 = 0, \label{trivoff}
\end{equation}
that gives
\begin{equation}
\frac{\dot A_1}{A_1} \frac{\dot A_2}{A_2} = \frac{\dot A_2}{A_2}
\frac{\dot A_3}{A_3} = \frac{\dot A_3}{A_3} \frac{\dot A_1}{A_1} = -
\vf_0^2 P_I. \label{trivoffnew}
\end{equation}
From \eqref{trivoffnew} one finds
\begin{equation}
\frac{\dot A_1}{A_1} = \frac{\dot A_2}{A_2} = \frac{\dot A_3}{A_3},
\label{rela123}
\end{equation}
that leads to the following relations between the three components
of vector potential:
\begin{equation}
A_1 = A, \quad A_2 = C_{21}A, \quad A_3 = C_{31}A,
\label{rela123new}
\end{equation}
with $C_{21}$ and $C_{31}$ being constants of integration.

In view of \eqref{rela123} and \eqref{trivoffnew} the diagonal
components of the energy momentum tensor take the form:
\begin{equation}
T_0^0 = - T_1^1 = -T_2^2 = -T_3^3 = \frac{\vf_0^2}{2} \Bigl[P + I
P_I\Bigr]e^{-2\alpha}. \label{temcom}
\end{equation}

Inserting \eqref{temcom} into \eqref{BID} for the metric functions
one finds:
\begin{subequations}
\begin{eqnarray}
\ddot \alpha - \dot \alpha^2 + \dot \beta_1^2 + \dot \beta_2^2 +
\dot \beta_3^2  &=& -\varkappa \vf_0^2 [P + I P_I], \label{00new1}\\
\ddot \beta_1 &=&  0, \label{11new1}\\
\ddot \beta_2 &=& 0, \label{22new1}\\
\ddot \beta_3 &=& 0.  \label{33new1}
\end{eqnarray}
\end{subequations}
From \eqref{11new1}, \eqref{22new1} and \eqref{33new1} we find
\begin{equation}
\beta_1 = b_1 t + \beta_{10}, \quad \beta_2 = b_2 t + \beta_{20},
\quad  \beta_3 = b_3 t + \beta_{30}. \label{betas}
\end{equation}
Here $b_i$ and $\beta_{i0}$ are integration constants. It should be
noted that in order to maintain the same scaling along all the axes,
the constants $\beta_{i0}$ should be the same, hence, without losing
generality one can set $\beta_{i0} = 0$.

Now let us go back to the electromagnetic field equations
\eqref{emf123}, which can be arranged as
\begin{subequations}
\label{emf123drov}
\begin{eqnarray}
\Bigl(\frac{\dot A_1}{A_1}\Bigr)^{\cdot} + \Bigl(\frac{\dot
A_1}{A_1}\Bigr)^2 - 2 \Bigl(\frac{\dot A_1}{A_1}\Bigr)\dot \beta_1 -
\vf_0^2 P_I &=& 0, \label{emf1drov}\\
\Bigl(\frac{\dot A_2}{A_2}\Bigr)^{\cdot} + \Bigl(\frac{\dot
A_2}{A_2}\Bigr)^2 - 2 \Bigl(\frac{\dot A_2}{A_2}\Bigr)\dot \beta_2 -
\vf_0^2 P_I &=& 0, \label{emf2drov}\\
\Bigl(\frac{\dot A_3}{A_3}\Bigr)^{\cdot} + \Bigl(\frac{\dot
A_3}{A_3}\Bigr)^2 - 2 \Bigl(\frac{\dot A_3}{A_3}\Bigr)\dot \beta_3 -
\vf_0^2 P_I &=& 0. \label{emf3drov}
\end{eqnarray}
\end{subequations}
In view of \eqref{rela123} from \eqref{emf123drov} we conclude that
\begin{equation}
\dot \beta_1 = \dot \beta_2 = \dot \beta_3, \label{eqdotbets}
\end{equation}
which is equivalent to $b_1 = b_2 = b_3 = b$ in \eqref{betas}, i.e.
\begin{equation}
\beta_1 = \beta_2 = \beta_3 = b t. \label{betasnew}
\end{equation}
As one sees from \eqref{betasnew}  we have isotropy at any given
time.

Now taking into account that both $A_2$ and $A_3$ can be expressed
in term of $A_1$ one could solve only one of the three equations of
\eqref{emf123}. In view of \eqref{rela123new} and \eqref{betasnew}
let us first rewrite \eqref{eminv} as follows
\begin{equation}
I =  - Q A^2 e^{-2bt}, \quad Q = [1 + C_{21}^2  + C_{31}^2 ]. \label{eminvnew}\\
\end{equation}

In view of \eqref{trivoffnew} and \eqref{betasnew} the equation for
$A$ now reads
\begin{equation}
A \ddot A + \dot A^2 - 2 b A \dot A  = 0. \label{eqA}
\end{equation}
One of the solutions to the equation \eqref{eqA} takes the form
\begin{equation}
A = D e^{bt}, \label{A}
\end{equation}
with $D$ being the constant of integration.

Let us now find the interaction corresponding to the solutions
obtained. Here we consider a few cases.

{\bf Case I} \qquad Let us assume $P$ be the power law of $I$: \quad
$P(I) = \lambda I^n$, where $\lambda$ is the coupling constant. For
$\lambda = 0$, we have ${\cG} = 1$. This corresponds to the
self-consistent system of scalar and electromagnetic fields with
minimal coupling. The off-diagonal component of the energy-momentum
tensor we write in the form
\begin{equation}
\dot A^2 + \vf_0^2 P_I A^2 = 0. \label{Pdet01}
\end{equation}
Inserting $A$ from \eqref{A} in this case we find
\begin{equation}
b^2 = - \lambda n \vf_0^2 I^{n-1}. \label{lambda01}
\end{equation}
From \eqref{lambda01} it becomes obvious that $n = 1$, that gives
\begin{equation}
b^2 = - \lambda \vf_0^2. \label{lambda001}
\end{equation}
From \eqref{lambda001} one concludes that $\lambda$ is negative.
Setting $\lambda = - \zeta$ one finds
\begin{equation}
b^2 =  \zeta \vf_0^2. \label{zeta001}
\end{equation}
 On the other hand from
\eqref{00new1} we find
\begin{equation}
3 b^2 = -\zeta \varkappa \vf_0^2 I. \label{zeta02}
\end{equation}
From \eqref{zeta001} and \eqref{zeta02} we find
\begin{equation}
I = -\frac{3}{\varkappa} = - D^2 (1 + C_{21}^2  + C_{31}^2),
\label{Il1}
\end{equation}
where the second equality follows from \eqref{eminvnew}.

Let us now once again go back to the electromagnetic field
equations, which in this case reads
\begin{equation}
\ddot A  + 2 b \dot A + b^2 A = 0, \label{emff}
\end{equation}
which is a Fock-Proca type equation. The equation \eqref{emff} shows
that the photon mass is directly related to the gravitational field.
In case of $b = 0$, i.e., in case of flat scape-time we have usual
Maxwell equation
\begin{equation}
\ddot A = 0. \label{max}
\end{equation}

{\bf Case II} \qquad Let us assume
\begin{equation}
{\cG} = \sum\limits_{n=0}^{\infty}\lambda_n I^n.\label{G}
\end{equation}
Recalling that ${\cG} = 1 + \Phi(I)$ one immediately finds
$\lambda_0 = 1$. Taking into account that $P = 1/G$ from
off-diagonal components of energy-momentum tensor we find
\begin{equation}
\varphi_0^2 G_I = b^2G^2.\label{od}
\end{equation}
In view of \eqref{od} equation  \eqref{00new1} now reads
\begin{equation}
(\frac{6}{\varkappa} + I)G_I = G. \label{00new2}
\end{equation}

The system \eqref{od} and \eqref{00new2} possesses a large number of
solutions. One of the simplest solution is
\begin{equation}
\lambda_1 = \frac{\varkappa }{6}, \qquad \lambda_2 \neq 0, \qquad
\lambda_{n>2}=0.\label{coef}
\end{equation}
In this case we get
\begin{equation}
I = -\frac{12}{\varkappa } =  - D^2 (1 + C_{21}^2  +
C_{31}^2).\label{rel10}
\end{equation}

{\bf Case with minimal coupling}

Let us now consider the case with minimal coupling. In this case we
have ${\cG} = 1/P = 1$. The purpose of this study is to clarify the
role of interaction in isotropization process. From the off-diagonal
components of energy-momentum tensor in this case we find
\begin{equation}
\dot A_1 \dot A_2 = \dot A_2 \dot A_3 = \dot A_3 \dot A_1 = 0.
\label{a123}
\end{equation}
From \eqref{a123} follows that at least two of the three components
$A_i$ are constant, which means only one of the components of
$F_{\mu\nu}$ is nonzero. Let us assume that $\dot A_1 = \dot A \ne
0$. In this case we have
\begin{subequations}
\label{emtemf}
\begin{eqnarray}
T_0^0 &=& \Bigl(\frac{\vf_0^2}{2} + \frac{1}{2} \dot A^2
e^{-2\beta_1}\Bigr)e^{- 2 \alpha}, \\
T_1^1 &=& \Bigl(-\frac{\vf_0^2}{2} + \frac{1}{2} \dot A^2
e^{-2\beta_1}\Bigr)e^{- 2 \alpha}, \\
T_2^2 &=& \Bigl(-\frac{\vf_0^2}{2} - \frac{1}{2} \dot A^2
e^{-2\beta_1}\Bigr)e^{- 2 \alpha}, \\
T_3^3 &=& \Bigl(-\frac{\vf_0^2}{2} - \frac{1}{2} \dot A^2
e^{-2\beta_1}\Bigr)e^{- 2 \alpha}.
\end{eqnarray}
\end{subequations}
In view of $\dot A_2 = \dot A_3 = 0$ from the electromagnetic field
equations in this case we have
\begin{equation}
A = C \int e^{2\beta_1} dt + C_1, \quad A_2 = {\rm const.}, \quad
A_3 = {\rm const.}, \label{emc}
\end{equation}
with $C$ and $C_1$ being some arbitrary constants. Einstein field
equations in this case take the form
\begin{subequations}
\begin{eqnarray}
\ddot \alpha - \dot \alpha^2 + \dot \beta_1^2 + \dot \beta_2^2 +
\dot \beta_3^2  &=& -\frac{\varkappa}{2} C^2 e^{2\beta_1} + \varkappa \vf_0^2, \label{00new10}\\
\ddot \beta_1 &=& -\frac{\varkappa}{2} C^2 e^{2\beta_1}, \label{11new10}\\
\ddot \beta_2 &=& \frac{\varkappa}{2} C^2 e^{2\beta_1}, \label{11new20}\\
\ddot \beta_3 &=& \frac{\varkappa}{2} C^2 e^{2\beta_1}.
\label{11new30}
\end{eqnarray}
\end{subequations}
In view of coordinate condition for $\alpha$ we find
\begin{equation}
\ddot \alpha = \frac{\varkappa}{2} C^2 e^{2\beta_1}.
\label{11newalpha}
\end{equation}

From \eqref{11new10}, \eqref{11new20} and \eqref{11new30}
immediately follows that in the case in question, no isotropization
process takes place. From \eqref{11new10} one finds the following
expression for $\beta_1$:
\begin{equation}
e^{2\beta_1} = \frac{2\eta^2}{\varkappa C^2} \frac{1}{{\rm cosh}^2
(\psi_0 - \eta t)}, \quad \eta^2 = {\rm const.}, \quad \psi_0 = {\rm
const.} \label{beta_10}
\end{equation}
Inserting $e^{2\beta_1}$ into \eqref{11new20}, \eqref{11new30} and
\eqref{11newalpha} one finds
\begin{equation}
e^{2\alpha} = e^{2\beta_2} = e^{2\beta_3} = \cosh^2{(\psi_0 - \eta
t)}. \label{ab2b3}
\end{equation}
From \eqref{beta_10} and \eqref{ab2b3} we get
\begin{equation}
\dot \alpha = - \dot \beta_1 = \dot \beta_2 = \dot \beta_3 = -\eta
{\rm tanh}(\psi_0 - \eta t). \label{abbbdot}
\end{equation}
Inserting \eqref{11newalpha}, \eqref{beta_10} and \eqref{abbbdot}
into \eqref{00new10} we find
\begin{equation}
\eta^2 =\frac{\varkappa \vf_0^2}{2}. \label{eta}
\end{equation}

Thus the system with minimal coupling is completely solved. It is
shown that in the case concerned, no isotropization process takes
place.

{\bf Case in absence of gravitational field}

Let us consider the case when the influence of the gravitational
field is not taken into account. In this case we have $\alpha =
\beta_1 = \beta_2 = \beta_3 = 0$, i.e., the space-time is flat. For
the scalar field in this case we have
\begin{equation}
{\dot \vf}^2 = \frac{\vf_0^2}{{\cG}^2}, \label{scf}
\end{equation}
whereas for the electromagnetic field we have
\begin{equation}
\ddot A + \vf_0^2 P_I A = 0. \label{elf}
\end{equation}
Unlike the case with gravitational field, now there is no
restrictions imposed on $P(I)$. If set, $P = \lambda I^n$, where $I
= A^2$ we now have
\begin{equation}
\ddot A + \vf_0^2 \lambda A^{2n -1} = 0. \label{elf1}
\end{equation}
In this case $n$ may take any value, with $n = 1$ giving the
Fock-Proca type equation.

\section{Conclusion}
Within the framework of Bianchi type-I cosmological model we studied
the evolution of the initially anisotropic space-time in presence of
an interacting system of electromagnetic and massless scalar fields.
We consider the case when the Lagrangian density of electromagnetic
field was given as a sum of Maxwellian part and the one explicitly
depending on scalar invariant ($\vf_{;\mu} \vf^{;\mu}$) and the
invariants $I= A_\mu A^\mu$, that was expressed as power law
function. It was shown that the model allows a set of partial
solutions, a few of which is described explicitly in this paper.

In case of interacting electromagnetic and scalar fields on account
of gravitational one, the Fock-Proca type equation with induced
massive-term was obtained. It was shown that only in case of
interacting material fields the isotropization process takes place.

It is shown that introduction of gravitational field, depending on
the concrete form of metric, imposes additional restriction to the
components of the vector potential.

In case of minimal coupling the isotropization process remains
absent.

\end{document}